\documentstyle[11pt,aaspp4,flushrt]{article}
\begin{document}
\newcommand{\xV}{\mbox{$\vec{x}$}}
\newcommand{\eff}{\rm{eff}}
\newcommand{\cl}{\rm{cl}}
\newcommand{\nV}{\mbox{$\vec{n}$}}
\newcommand{\kV}{\mbox{$\vec{k}$}}
\newcommand{\aV}{\mbox{$\vec{\alpha}$}}
\newcommand{\gV}{\mbox{$\vec{\gamma}$}}
\newcommand{\xiV}{\mbox{$\vec{\xi}$}}
\newcommand{\thV}{\mbox{$\vec{\theta}$}}
\newcommand{\bV}{\mbox{$\vec{\beta}$}}
\newcommand{\dV}{\mbox{$\vec{\nabla}$}}
\newcommand{\xip}{\mbox{$\vec{\xi}'$}}
\newcommand{\xP}{\vec{x}_{\perp}}
\newcommand{\dPL}{\vec{d}_{\perp}^{L}}
\newcommand{\xPL}{\vec{x}_{\perp}^{L}}
\newcommand{\xPS}{\mbox{$\vec{x}_{\perp}^{S}$}}
\newcommand{\nP}{\mbox{$\vec{n}_{\perp}$}}
\newcommand{\nPO}{\mbox{$\vec{n}_{\perp}^{0}$}}
\newcommand{\nPL}{\mbox{$\vec{n}_{\perp}^{L}$}}
\newcommand{\dP}{\vec{\nabla}_{\perp}}
\newcommand{\Hinv}{\mbox{$H_{0}^{-1}$}}
\newcommand{\tO}{\mbox{$\tau_{0}$}}
\newcommand{\tL}{\mbox{$\tau_{L}$}}
\newcommand{\tS}{\mbox{$\tau_{S}$}}
\newcommand{\tp}{\mbox{$\tau '$}}
\newcommand{\zL}{\mbox{$z_{L}$}}
\newcommand{\zS}{\mbox{$z_{S}$}}
\newcommand{\rL}{\mbox{$r_{L}$}}
\newcommand{\rS}{\mbox{$r_{S}$}}
\newcommand{\ddP}{\mbox{$\vec{d}_{\perp}$}}
\newcommand{\dPS}{\mbox{$\vec{d}_{\perp}^{S}$}}
\newcommand{\lP}{\mbox{$\vec{l}_{\perp}$}}
\newcommand{\lPO}{\mbox{$\vec{l}_{\perp}^{0}$}}
\newcommand{\pP}{\mbox{$\vec{p}_{\perp}$}}
\newcommand{\Hpres}{\mbox{$100h\,{\rm km\,sec^{-1}\,Mpc^{-1}}$}}

\title{
EFFECT OF LARGE-SCALE STRUCTURE ON MULTIPLY IMAGED SOURCES}

\author{Rennan Bar-Kana\footnote{email: barkana@arcturus.mit.edu}}
\affil{Department of Physics, MIT, Cambridge, MA 02139 USA}

\begin{abstract}
We study the effects of large-scale density fluctuations on
strong gravitational lensing. Previous studies have focused
mostly on weak lensing, since large-scale structure  alone 
cannot produce multiple images. When a galaxy or cluster
acts as a primary lens, however, we find that large-scale
structure can produce asymmetric shear of the same order as 
the lens itself. Indeed, this may explain the origin of the 
large shear found in lens models in conflict with the small 
ellipticity of the observed galaxy light distributions. We 
show that large-scale structure changes the lens equation 
to the form of a generalized quadrupole lens, which affects 
lens reconstruction. Large-scale structure also changes the 
angular diameter distance at a given redshift. The precise 
value depends on the lens and source redshifts and on the 
large-scale structure power spectrum, but the induced
$1\sigma$ uncertainty in determinations of the Hubble constant 
from measurements of time delays is of order $5-10 \%$. If 
observations of lensing can constrain the magnitude of the
shear which is due to large-scale structure, it would provide 
a direct probe of the overall amplitude of mass fluctuations. 
\end{abstract}


\vspace{.2in}

\section{Introduction}

Gravitational lensing is one of the most promising methods
of mapping the distribution of matter at cosmological distances.
Detailed observations of multiple images of quasars have been
used to try and reconstruct the lensing mass distribution (e.g.\
\cite{falco}). It has also long been recognized that
measurements of the time delay between images can be used to
determine the Hubble constant (\cite{refsdal64}, 1966).
However, practical application to the double quasar 0957+561
has been difficult because of uncertainties in lens modelling
as well as conflicting measurements of the time delay
(e.g.\ \cite{td1}; \cite{td2}).

Since gravitational lenses and sources typically lie at significant
redshifts, light rays are deflected by large-scale structure
(LSS) as they traverse the enormous distance from source to observer.
These deflections are not large enough to produce multiple images,
but they do distort the shapes of sources. Such weak lensing
has been investigated both analytically (e.g.\ \cite{miralda}; 
\cite{kaiser}) and in N-body simulations (\cite{jaros};
\cite{blandford}). These studies find a shear of order $1\%$,
coherent over a scale of $\approx 1^{\circ}$, in a flat CDM
model. This shear may in principle be detected observationally as a
coherent distortion of background galaxies, when averaged
over a sufficiently large angular field in order to be
separated from the random scatter of intrinsic ellipticities 
(e.g.\ \cite{mould}; \cite{vill}). 

The shear due to LSS can also affect strong lensing, when it
acts in addition to a strong primary lens, a galaxy or cluster
near the line of sight to the source. This effect is enhanced
compared to weak lensing, because of the small angular separations
between multiple images. Also, the higher redshift 
of quasars compared to faint galaxies increases the cumulative
shear from the observer to the source. Seljak (1994) 
estimated the dependence of the r.m.s.\ value of this shear
on the power spectrum of density fluctuations, and found
it to be of order $10 \%$ for a source at redshift $3$.
Seljak also considered the effect of LSS on the time delay,
and showed that the lowest order terms cancel out in the total
time delay. However, since these canceling terms are separately
much larger than the time delay from the primary lens, even
higher order terms might still dominate the time delay and
threaten the effort to determine the Hubble constant from lensing.

In order to find precisely how LSS affects the observables of a
lens system, Surpi et al.\ (1995) set up the lens equation in
the presence of a lens plus LSS. They made an expansion for
the position of a light ray in powers of its deflection from the 
unperturbed straight path, and kept only the lowest order term.
This term is equivalent to a constant angular deflection at the
lens. They thus concluded that LSS leaves 
all observables (such as relative image positions) unchanged to 
lowest order. Indeed, since the actual source position is unobservable, 
the effect of this lowest order term can be removed from the lens 
equation by subtracting the constant angle out of the source angle.
This approximation of keeping the lowest order term is not
a good one, however, since the shear due to LSS arises from
{\it relative} deflections between different light rays, which
involve higher order terms in the expansion. We follow a
similar approach but include these higher order
terms in order to study the observable effects of LSS.

In this paper we analyze the effect of LSS on the lens equation 
and time delay. Readers primarily interested in the results may
wish to concentrate on \S 4 -- \S 6 and \S 8. We derive the lens 
equation in \S 2 and \S 3, and find it to have a form similar to the 
generalized quadrupole lens of Kovner (1987) (\S 4).
We express the perturbed lens equation in terms of integrals along the
line of sight of the scalar, Newtonian potential.
These integrals are random variables of zero mean, whose variances 
and covariances can be evaluated in terms of the
power spectrum of density perturbations (\S 5). We find that
the effective shear in the lens equation is not simply the integrated
shear from the observer to the source, but is reduced by $40 \%$
or more, depending on the lens redshift. For realistic power spectra 
that include modelling of non-linear effects, the effective shear is 
of order $6 \%$ for a source at redshift $3$. In addition, the accumulated
shear from the observer to the lens can significantly affect the
observables as well as the appearance of the lens itself, if the lens is 
at a relatively high redshift. In \S 3 we also discuss how our results 
determine the effect of LSS on angular diameter distances.

The most important effect of shear is in producing four-image systems.
Many confirmed lens systems are quads, since they are easy to identify
and tend to be highly magnified (Kochanek 1991b,1995; Wallington 
\& Narayan 1993). These systems are inconsistent with
an axi-symmetric lens, for which all the images would have to be
colinear. Lens models of quads typically find a shear of order $7-11 \%$
(e.g.\ \cite{koch1}).
If due to the lensing galaxy itself, this would imply a projected ellipticity
($=1$ minus the ratio of minor to major axis)
for the mass of $\approx 35-50 \%$. By contrast, the typical value observed
for ellipticals is $\approx 20 \%$ (e.g.\ Ryden 1992; Schechter 1987). 
Since the cross-section for producing quads increases with shear,
observed quads should be biased towards high shear, whatever its origin
(\cite{kass}). In particular this includes a bias toward an alignment 
between the shear caused by the galaxy and the external shear.
High resolution observations of lensing galaxies can determine
the degree of agreement or inconsistency between the observed ellipticity 
and the inferred shear in specific cases. Recent observations of
a four-image ``Einstein cross'' with the Hubble Space Telescope WFPC2 
(\cite{hst1}) found an ellipticity in the potential of $26 \%$, which
implies a mass ellipticity of $60 \%$. The light distribution was 
found to have an ellipticity of only $32 \%$. One possible explanation is 
that the dark matter halo is highly flattened compared to the light
distribution, but other observations may not support the existence of
such large differences in typical galaxies (for a recent review see 
\cite{sack}). Another possibility is that a LSS shear of order $8 \%$ has 
been added on to the $7 \%$ shear of the lens. In fact, the directions
of the total shear and that due to the light distributions are different 
by about $13^{\circ}$, so the disagreement is larger. A recent HST observation 
of a lensed arc (\cite{hst2}) has similarly found an observed ellipticity of 
about half that implied by the best fit lens model. Note, however
that other possible sources of external shear, namely additional
galaxies or clusters near the line of sight to the source, must
be properly accounted for before the contribution of LSS can be determined.

In \S 4 we also consider the effect of LSS on relative time delays
of images. The related phenomenon of amplification of sources due to
large-scale structure has been studied by Babul \& Lee (1991), but
not in the presence of a primary lens. We show that the effect on
time delays is enhanced through a combination of two separate effects. 
LSS thus limits our ability to derive accurate values of the Hubble 
constant from lensing. The induced uncertainty depends on the
lens and source redshifts and on the large-scale structure
power spectrum, but in \S 5 we find it to be of order $5-10\%$ at $1\sigma$.
This uncertainty may have either sign since LSS may effectively produce a 
negative mass density (negative is measured w.r.t.\ the mean density
of the universe, not w.r.t.\ zero).

In \S 6 we choose a simple lens distribution, a singular isothermal sphere, 
and illustrate the effect of LSS on relative image positions and time 
delays, as well as the caustics and critical curves of the lens system.
In \S 7 we apply our formalism to the transition from strong to weak 
lensing and demonstrate its agreement with previous studies of weak lensing
(a detailed derivation is given in Appendix B).
Finally, in \S 8 we give our conclusions and comment on possible
applications of our results.

We assume a flat universe throughout, in the absence of an accurate 
fitting formula for the time dependent, non-linear power spectrum
in a curved background. Our formalism is, however, easily generalized
to a closed or open universe, as we show in Appendix A.

\section{Formalism}

We work in the framework of a flat Robertson-Walker metric
with small-amplitude scalar metric fluctuations. In the
longitudinal gauge (\cite{bard80}) we can write the
line element as
\begin{equation}
ds^2=a^2(\tau)[-(1+2\phi)d\tau^2+(1-2\phi)
d\xV \cdot d\xV~]\ ,
\label{metric}
\end{equation}
where we set ${\rm c}=1$. Here $\tau$ is the conformal time,
$a(\tau)$ the expansion factor, and we are using comoving
coordinates $\xV$. Redshift in a flat, matter-dominated
($\Omega_{m}=1$) universe is given by $1/(1+z)=a(\tau)=
(H_{0}\tau/2)^2$, 
with $H_{0}=\Hpres$ the present Hubble constant.
Also $\phi$ is the scalar, Newtonian
potential obeying the cosmological Poisson equation
\begin{equation}
\nabla^2\phi=4 \pi G a^{2} \bar{\rho}\ \delta\ , 
\label{Pois}
\end{equation}
where $\bar{\rho}$ is the mean density of the universe
and $\delta=(\rho-\bar{\rho})/\bar{\rho}$ is the local density 
perturbation. We describe statistical properties of $\phi$ in terms
of its Fourier transform $\phi(\kV,\tau)$, where
$\phi(\vec x,\tau)=\int d^3k \phi(\vec k,\tau)e^{i\vec k \cdot 
\vec x}$. Its ensemble mean and variance
are $\langle \phi(\vec k,\tau)\rangle=0$
and $\langle \phi(\vec k,\tau)\phi^*(\vec k',\tau)\rangle=
P_{\phi}(k,\tau)  \delta^3(\vec
k-\vec k')$, where $P_{\phi}(k,\tau)$ is the power spectrum 
of the potential at time $\tau$, simply related to the density 
power spectrum by $P_{\phi}(k,\tau) = \left ( 4 \pi G a^{2}(\tau)
\bar{\rho}(\tau) \right )^{2} k^{-4} P_{\rho}(k,\tau)$ .

We place the observer at the origin of coordinates and the
primary lens \footnote{``The lens'' refers to some
reference point in the lens plane, such as the center
if the lens is axi-symmetric.} 
on the $z$-axis. We use $r$ to denote values of
the $z$-coordinate, with $\zL$ and $\zS$ reserved for
lens and source redshift, respectively. Note that the $z$-axis
is only a coordinate axis used for reference and not the
actual path of any light ray. 
We let $\vec{n}$ denote a unit vector in the photon's direction of 
motion and $\vec{x}$ denote its position. To first order in $\phi$,
in the metric (\ref{metric}) they obey 
\begin{equation}
\frac{d\nV}{d\tau}=-2 \left[\dV\phi - \nV (\nV\cdot\dV\phi)\right]\ , \ \ \ \ 
\ \frac{d\xV}{d\tau}=\nV(1+2\phi)\ .
\label{roftau}
\end{equation}
In this and similar expressions in this section, $\phi$ is to be 
evaluated on the actual photon path, not on the $z$-axis.

We now assume that the angle between $\nV$ and the $z$-axis is small
(e.g.\ \cite{seljak}), and consider the components
perpendicular to the $z$-axis of $\nV$ and $\xV$. They obey
\begin{equation}
\frac{d\nP}{d\tau}=-2 \dP \phi\ , \ \ \ \ \ 
\frac{d\xP}{d\tau}=\nP\ ,
\label{eqofmot}
\end{equation}
where $\dP \phi$ denotes the derivative of the potential transverse 
to the $z$-axis. In the approximation of small angles, these equations are the 
same as the Newtonian equations of motion for a particle moving in a 
gravitational field, except for the factor of $2$ from General Relativity.
The absolute mean of $\phi$ is not observable,
since the perturbations in the metric are defined about the large-scale mean.
Indeed, we may choose our space and time units so that the large-scale value
of $\phi$ is zero at the Local Group. Then $\phi$ is a random variable 
with r.m.s.\ value of order $10^{-4}$ for the observed LSS power spectrum.
Equation (\ref{roftau}) implies that the photon path obeys $r(\tau)=\tO-\tau$ with 
$O(\phi)$ corrections, where 
$\tO$ is the present value of $\tau$. The relation of comoving distance to redshift 
is, e.g.\ in an Einstein-de Sitter universe, $r(z)=2 \Hinv [1-(1+z)^{-1/2}]$.
Thus comoving distances are simply related to the measured redshifts 
(to $O(\phi)$), and so we use comoving distances rather than angular diameter
distances ($\rL$ and $\rS$ refer below to the lens and source, respectively). 
In a homogeneous universe with no LSS, angular diameter distances 
are given by $D=r/(1+z)$. In general, if an object at comoving distance 
$r$ has a proper diameter $R$ and is observed to subtend an angle
$\theta$, then the angular diameter distance is defined to be 
$D=R/\theta$. This differs from $D$ in the homogeneous case by terms larger
than $O(\phi)$. As explained in \S 1, this is precisely the effect which we 
calculate below, and so we discuss angular diameter distances further in \S 3.

We can trace the photon trajectory backwards in time using equations 
(\ref{eqofmot}), with the final conditions $\xP=0$ and $\nP=\nPO$ at
the observer $r=0$. Between the observer and the lens,
we find that
\begin{eqnarray}
\nP(\tau)&=&\nPO+2\int_{\tau}^{\tau_{0}}\dP\phi(\tp)d\tp\ , 
\nonumber \\
\xP(\tau)&=&-(\tO-\tau)\nPO -2\int_{\tau}^{\tau_{0}}
(\tp-\tau)\dP\phi(\tp)d\tp\ .
\label{eq1}
\end{eqnarray}
When the photon is at the lens, its direction of motion is
$\nP(\tL)$. It is then deflected so that at the source side of the 
lens its direction of motion is $\pP=\nP(\tL)+\gV$. The 
deflection angle $\gV$ is evaluated at $\xPL \equiv \xP(\tL)$, and is 
determined by the mass distribution of the primary lens.
Equations (\ref{eqofmot}) then imply that, between the lens and
the source, 
\begin{eqnarray}
\nP(\tau)&=&\pP+2\int_{\tau}^{\tau_{L}}\dP\phi(\tp)d\tp\ ,
\nonumber \\
\xP(\tau)&=&\xP(\tL)-(\tL-\tau)\pP-2\int_{\tau}^{\tau_{L}}(\tp-\tau)
\dP\phi(\tp)d\tp\ .
\label{eq4}
\end{eqnarray}
For a given source position $\xPS$, the lens equation is then $\xP(\tS)=\xPS$.

The total proper time delay, relative to the $\phi=0$ path 
along the $z$-axis, is given by (e.g.\ \cite{glbook}) 
\begin{eqnarray}
\Delta t = \int_{\tau_{S}}^{\tau_{0}}\left[\frac{1}{2}\left(\frac{d\xP}
{d\tau}\right)^{2}-2\phi\right]\/d\tau~
-(1+\zL)\psi(\xPL)\ . 
\label{tdgeneral}
\end{eqnarray}
The first term is the geometrical time delay, the second is the
potential time delay due to LSS, and the last is the potential
time delay due to the lens, given by
\begin{equation}
\psi(\xP) = 4G\int d^{2}\xi ' \Sigma(\xip)~\log\left|\frac{
\xiV-\xip}{(1+z_{L})^{-1}\rL}\right|\ ,
\end{equation}
where $\Sigma(\xiV)$ is the projected mass density of the lens,
and $\xiV=(1+z_{L})^{-1}\xP$ measures {\it proper} distance in the lens 
plane. We let $r_{LS}=r_{S}-r_{L}$, and then the scaled deflection angle is 
given by 
\begin{equation}
\aV = \frac{r_{LS}}{r_{S}} \gV =
\frac{r_{LS}}{r_{S}}\frac{\partial \psi}{\partial \xiV}\ .
\end{equation}

\section{Lensing in the presence of LSS}

Equations (\ref{eq1}) and (\ref{eq4}) cannot in 
general be solved analytically, since they
involve integrals over the potential $\phi$
evaluated on the (unknown) photon path.
We therefore expand $\phi$ about its value on the
$z$-axis (as in \cite{surpi}):
\begin{equation}
\phi(r \hat{z} + \xP) \approx \phi + \xP \cdot \dP \phi\ +
\frac{1}{2}(\xP \cdot \dP)^{2} \phi\ ,
\label{phi}
\end{equation}
where the right-hand side (RHS) is evaluated on the $z$-axis. 
The second term on the RHS leads to an
unobservable constant deflection, and the third
to a relative deflection between light rays.
Unlike Surpi et al., we include the third term. 
To lowest order, in the resulting expansion for $\dP \phi$ we substitute 
for $\xP$ the expressions given in equations (\ref{eq1}) and (\ref{eq4}) 
with $\phi$ evaluated on the $z$-axis. Hereafter,
all expressions involving $\phi$ are evaluated on the $z$-axis.

The expansion (\ref{phi}) should be valid as long as the LSS power on scales
smaller than the deflection $\xP$ is negligible. We find below, however,
that the shear is produced by modes over a broad range of wavelengths.
Moreover, the higher order terms in the expansion formally diverge at small 
wavelengths in an r.m.s.\ sense, e.g.\ for a scale-invariant spectrum, at fixed 
$\xP$. In reality, $\xP$ depends on the initial direction and on $\phi$. 
This worry is resolved
by using a different expansion, equivalent to summing this entire series
(see \S 7). This alternate expansion is convergent, and shows that the 
contribution of small wavelength modes is cut off. For strong lensing
we find that the terms in equation (\ref{phi}) suffice for an
accurate analysis. Note that we have not assumed at any point that $\delta < 1$
for the density. Our expansions remain valid even when we include non-linear
modes for which $\delta \gg 1$. 

\begin{figure}[t]
\vspace*{6.3 cm}
\caption{Sketch showing the fiducial ray and an image ray,
with distances in comoving coordinates.}
   \includegraphics{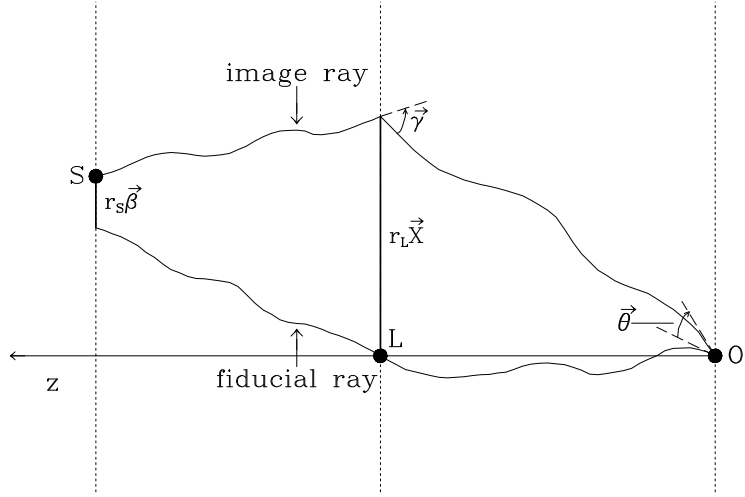}
\label{sketch}
\end{figure}

We are not interested in any deflection which is common to all light rays,
since such a constant angle only affects the unobservable absolute
position of the source. We can subtract out such terms to all
orders simply by measuring displacements relative to some light ray
instead of the $z$-axis.
We define this fiducial ray as the light ray (null geodesic) passing
through the observer and through the lens, and
extended out to $\rS$ (see figure \ref{sketch}). This ray is 
deflected by LSS throughout its path, but is not deflected
by the primary lens. The quantities $\xP$ and $\nP$
measured relative to the corresponding quantities for
this fiducial ray we denote by $\ddP$ and $\lP$,
respectively. Then $\thV=-\lPO$ is the observed
angle of a light ray relative to the observed
lens position. Note that $\dPL=\xPL \equiv r_{L}\vec{X}$.

We define dimensionless $2 \times 2$ symmetric tensors,
\begin{eqnarray}
{\rm F}_{ij}(\tau_{1},\tau_{2}) &=& -\frac{2}{\tau_{1}-\tau_{2}}\int
_{\tau_{2}}^{\tau_{1}}(\tau-\tau_{2}) 
(\tau_{1}-\tau)\partial_{i}\partial_{j}\phi(\tau)d\tau\ , \nonumber \\
{\rm G}_{ij}(\tau_{1},\tau_{2}) &=& -2 \int
_{\tau_{2}}^{\tau_{1}}(\tau_{1}-\tau)\partial_{i}
\partial_{j}\phi(\tau)d\tau\ .
\end{eqnarray}
The traceless part of F$_{ij}$ is the shear 
tensor of weak lensing (e.g.\ \cite{kaiser}). 

Between the observer and the lens,
\begin{eqnarray}
l_{\perp}^{i}(\tau)&=&-\theta^{i}-\theta_{j}{\rm G}^{ij}(\tO,\tau)
\ , \nonumber \\
d_{\perp}^{i}(\tau)&=&(\tO-\tau)\left[\theta^{i}+
\theta_{j}{\rm F}^{ij}(\tO,\tau)\right]\ .
\label{thiseq}
\end{eqnarray}
Equations (\ref{thiseq}) suggest a simple physical interpretation for the
two tensors, in our approximation. For two rays that end up at the
origin at $\tau_{0}$ with a small angular separation
$\thV$, $(\tau_{0}-\tau){\rm F}_{ij}\theta^{j}\ $ \footnote{Repeated 
indices are summed over the x and y directions. There is no distinction
between upper and lower indices.}
measures the change in their relative separation at $\tau$, compared
to having no LSS.
G$_{ij}\theta^{j}$ similarly measures the induced change in their relative
directions at time $\tau$. 

If we let $\pP=\lP(\tL)+\gV\left(\dPL\right)$ then, between the
lens and the source, 
\begin{eqnarray}
l_{\perp}^{i}(\tau)&=&p_{\perp}^{i}+p_{\perp}^{j}{\rm G}^{i}_{j}(\tL,\tau)- 
d_{\perp}^{j}(\tL)\left[{\rm G}^{i}_{j}(\tL,\tau)+{\rm G}^{i}_{j}(\tau,\tL)
\right]/(\tL-\tau)\ , \nonumber \\
d_{\perp}^{i}(\tau)&=&d_{\perp}^{i}(\tL)-(\tL-\tau)\left[p_{\perp}^{i}+
p_{\perp}^{j}{\rm F}^{i}_{j}(\tL,\tau)\right] 
 +d_{\perp}^{j}(\tL){\rm G}^{i}_{j}(\tau,\tL)\ . 
\end{eqnarray}
The lens equation is $\ddP(\tS)=\dPS$.

Defining $\bV=\dPS/r_{S}$, and denoting e.g.\ F$^{ij}(\tO,\tL)$ by
F$_{OL}^{ij}$, the lens equation becomes
\begin{equation}
\beta^{i}=\theta^{i}+\theta_{j}{\rm F}_{OS}^{ij}-(\alpha^{i}
+\alpha_{j}{\rm F}_{LS}^{ij})\ ,
\label{lens}
\end{equation}
where $\aV$ is evaluated at 
\begin{equation}
d_{\perp}^{i}(\tL)=\rL(\theta^{i}+
\theta_{j}{\rm F}^{ij}_{OL})\ .
\label{dperp}
\end{equation} 
We thus conclude that to our order of approximation, LSS
affects the lens equation through three terms, which are
easily understood. Two rays separated by an
angle $\theta^{i}$ at the observer would, in the absence of lensing
or LSS, 
be separated by a (comoving) distance $r_{L} \theta^{i}$ at the lens 
and $r_{S} \theta^{i}$ at the source. The LSS shear changes these
distances to $r_{L} (\theta^{i}+\theta_{j}{\rm F}^{ij}_{OL})$
and $r_{S} (\theta^{i}+\theta_{j}{\rm F}_{OS}^{ij})$ respectively.
The deflection between the two rays by an angle
$-\gamma^{i}$ at the lens leads to an additional separation
of $-r_{LS} \gamma^{i}=-r_{S}\alpha^{i}$ at the source, or $-r_{S} (\alpha^{i}
+\alpha_{j}{\rm F}_{LS}^{ij})$ when we include the effect of LSS
shear. There are no cross-terms between these three effects 
in our approximation, where only terms linear in $\thV$
and $\aV$ appear in $\ddP$.

The magnification matrix is
\begin{equation}
\frac{\partial\beta^{i}}{\partial\theta_{j}}=\delta^{ij}
-(-{\rm F}_{OS}^{ij}+\Psi^{ij}+\Psi^{i}_{k}{\rm F}_{OL}^{kj}
+\Psi^{j}_{k}{\rm F}_{LS}^{ki})\ ,
\label{eqmag}
\end{equation}
where $\Psi^{ij}=\partial_{\xi_{i}}\partial_{\xi_{j}}\psi/
(4 \pi G \Sigma_{cr})$ is the
shear matrix of the primary lens, in units of $\Sigma_{cr}=
r_{S}(1+z_{L})/(4 \pi G r_{L} r_{LS})$, and $\Psi^{ij}$
in equation (\ref{eqmag}) is evaluated at $\dPL$.
With the usual sign conventions in lensing, the constant LSS 
shear is $-{\rm F}_{OS}^{ij}$. This term would still be 
present even in the absence of the lens (see \S 7 below).
Note that $\partial\beta^{i}/\partial\theta_{j}$ is
in general {\it not} symmetric, which it would be in the
absence of LSS. In other words, LSS can rotate images.

As noted in \S 1, we have also calculated the effect of LSS on angular
diameter distances. Indeed, an object which subtends an angle $\thV$
at the observer measures a comoving distance $\dPL$ on the lens plane,
given by equation (\ref{dperp}). The same object measures a proper
distance $\vec{R}=\dPL/(1+\zL)$, which follows (to $O(\phi)$) from the line 
element (\ref{metric}) taken at constant $\tau$. Then $R^i=D_{OL}^{ij}\theta_j$,
where the angular diameter ``distance'' $D_{OL}^{ij}=(\delta^{ij}+F_{OL}^{ij})r_L
/(1+z_L)$ is a tensor when LSS is present. Thus $R$ at a given $\theta$ depends on 
orientation, and also $\vec{R}$ may have a different direction than $\thV$,
so when giving ``distances'' to lenses it is preferable to use the comoving distance 
$\rL$ which is well defined (up to corrections of $O(\phi)$, i.e.\ $0.01\%$) in 
terms of $\zL$. As we show in \S 5, the components of F$_{OL}$ are of order a few 
percent, much larger than $O(\phi)$ corrections. 
Some other authors (e.g.\ Ehlers \& Schneider 1986, Watanabe et al.\ 1992, Sasaki 
1993) have also considered the effect of LSS on angular diameter distances, but 
they used an oversimplified model in which some fraction of the mass density in the 
universe is distributed in clumps. Theory and observation of LSS indicate that a 
description in terms of a random field with positive and negative fluctuations over a 
range of scales is more realistic (see also the related discussion in \cite{seljak}).

\section{The Lens Equation and Time Delay}

The lens equation (\ref{lens}) is similar in form to the
generalized quadrupole lens of Kovner (1987). Kovner
cosidered multiple lensing in which there is one primary
lens and additional lenses with linear deflection
laws. In that case Kovner showed how to write the
lens equation in the form of a thin-lens equation, which
simplifies the analysis of properties of the lens mapping, 
such as image multiplicities and the time delay between
images. LSS is different in that the deflection
is accumulated continuously, but the final result can
be similarly simplified.
Letting 
\begin{eqnarray}
\label{xtheta}
X^{i} & = & \theta^{i}+\theta_{j}{\rm F}_{OL}^{ij}\ , \\
\label{ytheta}
Y^{i} & = & \beta^{i}-\beta_{j}{\rm F}_{LS}^{ij}\ , 
\end{eqnarray}
the lens equation becomes
\begin{eqnarray}
\label{lenskov}
Y^{i}=X^{i}-\left[X_{j}{\rm F}_{\eff}^{ij}+\alpha^{i}(\vec{X})\right]\ , \\
{\rm F}_{\eff}^{ij}=-{\rm F}_{OS}^{ij}+{\rm F}_{LS}^{ij}+{\rm F}_{OL}^{ij}\ ,
\end{eqnarray}
where we write $\aV$ as a function of $\vec{X}$ rather than
$r_{L}\vec{X}$. We find that $\delta^{ij}-{\rm F}_{\eff}^{ij}$
plays the same role as the ``telescope matrix'' of Kovner, which in
his case is in general symmetric. We find a symmetric F$_{\eff}$ only 
because we are working to first order in the LSS shear. The effective 
shear F$_{\eff}^{ij}$ is in general significantly weaker than F$_{OS}^{ij}$,
as we show in \S 5. Still, this shear should be of order $6 \%$ r.m.s.,
compared e.g.\ with a galaxy of ellipticity $20 \%$, which produces
a shear of $\approx 4 \%$. 

We thus have a simple description of the lens mapping: The source
plane is slightly distorted to become the $\vec{Y}$ plane, as given
by equation (\ref{ytheta}), so e.g.\ a circular source appears
elliptical in the $\vec{Y}$ plane. Equation 
(\ref{lenskov}) then gives the lens mapping from the $\vec{Y}$ plane
to the $\vec{X}$ plane. Finally, the (observed) image plane is also
a slightly distorted picture of the $\vec{X}$ plane, as in
equation (\ref{xtheta}). Only the $\vec{Y} \mapsto \vec{X}$
map is non-linear, and it determines the geometry of the lens
mapping. Thus e.g.\ the probability of producing quads depends on the
sum of F$_{\eff}$ and any intrinsic asymmetric shear from the ellipticity
of the lens galaxy. The shear F$_{\eff}$ should tend to make the observed 
galaxy light distribution inconsistent with the observed lensing. 
If the lens is at high redshift, however, then the distortion in equation 
(\ref{xtheta}) is also important, since $\thV$ is observed and not $\vec{X}$.
In this case, the lens itself is distorted by LSS, if it is observed. Because 
this induced ellipticity is likely to be wrongly interpreted as 
intrinsic to the galaxy, it tends to confuse observers as to the actual direction 
of the galaxy's internal shear, but the effect is important only if the intrinsic
ellipticity itself is not too large.
Since the source plane is not directly observable, the distortion in equation
(\ref{ytheta}) does not affect lens reconstruction, but it is
important for absolute magnifications (given by equation 
(\ref{eqmag})), and for measuring shape distortions (\S 7).

We can calculate the time delay explicitly using equation (\ref{tdgeneral}).
However, it is easier to use Fermat's principle, which implies that
(for a given $\phi(\vec{x},\tau)$)
the lens equation must be equivalent to $\partial\Delta t/
\partial\thV=0$ at fixed $\bV$ (e.g.\ \cite{glbook}). Thus
the time delay is the same as that corresponding to the thin-lens 
equation (\ref{lenskov}) which, up to $\vec{X}$-independent terms, 
equals (Kovner 1987)
\begin{equation}
\Delta t = \frac{1}{2} \frac{\rL\rS}{r_{LS}} \left[ (\vec{X}-\vec{Y})^{2}
-\mbox{F}_{\eff}^{ij}X_{i}X_{j} \right]
-(1+\zL)\psi(r_{L}\vec{X})\ .
\label{tdelay1}
\end{equation}
We might have expected linear terms of the form $\theta^{i}{\rm 
C}_{i}$ to make large contributions to $\Delta t$, where C$_{i}$ is 
independent of $\thV$ and $\bV$, e.g.\ C$_{i}=r_{LS}\int_{\tau_{S}}^
{\tau_{L}}\partial_{i}\phi(\tau)d\tau$ or C$^{i}={\rm F}_{LS}^{ij}
\int_{\tau_{S}}^{\tau_{L}}(\tL-\tau)\partial_{j}\phi(\tau)d\tau$.  
Such terms do appear in the geometric and potential time delays,
but Fermat's principle shows that they must
drop out in the total time delay. These cancellations can also be 
demonstrated explicitly with equation (\ref{tdgeneral}).

In addition to shear effects, the lens geometry is also affected
by the trace parts of F$_{\eff}$ and F$_{OL}$.
These traces cannot be determined through lens reconstruction,
since they only affect the unobservable overall scales of the
lens size and distance. However, they do affect the determination
of the Hubble constant from lensing. To show the various effects, we first
set F$_{OL}^{ij}=0$, and consider just equation (\ref{lenskov}) and the 
effect of the trace part of F$_{\eff}$. We also allow for a focusing term 
$\kappa_{cl}$ due to a cluster surrounding the lens galaxy. Equation $5$ of 
Narayan (1991) applies here:
\begin{equation}
\Delta t \propto \sigma^2 r_L\ ,
\label{delt1}
\end{equation}
where $\sigma$ represents some characteristic velocity of the lens system.
The proportionality constant in this equation depends on a number of parameters which 
can in principle be determined for each pair of images from lens modelling.  
One of these parameters involves $\sigma$ and $\kappa_{\cl}$, in the
combination
\begin{equation}
\zeta=\frac{r_S}{\sigma^2 r_{LS}}\left[1-\left(\frac{1}{2}{\rm Tr\ F}_{\eff}
+\kappa_{\cl}\right)\right]
\label{zeta}
\end{equation}
(as follows from equation $6$ of Narayan (1991)).
Thus if $\Delta t$ is measured (and $\sigma$ is not) then from the product 
$\zeta \Delta t$, we can try to deduce $H_0$ given $z_L,\ z_S$, and an assumed 
deceleration parameter $q_0$. The real $H_0$ is different from the $H_0$
deduced assuming $\kappa_{\cl}={\rm Tr\ F}_{\eff}=0$ by a factor
of $[1-(\frac{1}{2}{\rm Tr\ F}_{\eff}+\kappa_{\cl})]$. 
If both $\Delta t$ and $\sigma$ can be measured independently, then as noted in 
Narayan (1991) we can (if F$_{OL}^{ij}=0$) circumvent the unknowns in equation 
(\ref{zeta}) and use equation (\ref{delt1}) to obtain $r_L$ directly, and thus 
$H_0$ for an assumed $q_0$. 

We now add in also the effect of the trace part of F$_{OL}$ in equation (\ref{xtheta}), 
which is an unobservable magnification of the lens plane produced by foreground 
structure. Since the time delay is proportional to the square of the angular scale 
(e.g.\ \cite{glbook}), equation (\ref{delt1}) is replaced by
\begin{equation}
\Delta t \propto \sigma^2 r_L \left(1+\frac{1}{2}{\rm Tr\ F}_{OL}\right)^2\ .
\label{delt2}
\end{equation}
Reasoning as above, we see that if only $\Delta t$ is measured then (from equations 
(\ref{zeta}) and (\ref{delt2})) the real global $H_0$ is different from the
deduced $H_0$ (for an assumed $q_0$) by a factor of $[1+\frac{1}{2}{\rm Tr}(2{\rm F}
_{OL}-{\rm F}_{\eff})-\kappa_{\cl}]$, to linear order. LSS thus 
produces a $1\sigma$ uncertainty in determinations of $H_0$ of
\begin{equation}
\sigma_{H_0,1}={\rm\ r.m.s.\ of }\ \frac{1}{2}{\rm\ Tr\ }(2{\rm F}_{OL}-
{\rm F}_{\eff})\ .
\end{equation}
Contrary to \cite{falco1}, we cannot derive an upper bound on $H_0$ from the $\Delta t$ 
measurement since while $\kappa_{\cl} \ge 0$, the LSS term may be negative or
positive. Even with measurements of both $\Delta t$ and $\sigma^2$, 
we cannot measure $H_0$ exactly, since when we use equation (\ref{delt2}) we
are subject (for a given $q_0$) to a $1\sigma$ uncertainty of
\begin{equation}
\sigma_{H_0,2}={\rm\ r.m.s.\ of }{\rm\ Tr\ }{\rm F}_{OL}\ .
\end{equation}
Thus, LSS creates uncertainties in determinations of the Hubble constant
from lensing which apply even to perfect lens models determined by an
arbitrarily large number of observables. If a precise measurement of
$H_0$ is sought from lens time delays, then at least for some redshifts
and LSS power spectra these uncertainties may not be very small, as we show
in \S 5.

\section{LSS effects in realistic models}

We have shown above that the effects of LSS on lensing enter
through the symmetric tensors F$_{OL}$,  F$_{LS}$ and F$_{OS}$.
For a given lens and source, these tensors will affect the
lens mapping as we showed in \S 4, possibly with observable
effects which we illustrate in \S 6. In this section we estimate
the typical magnitude of these tensors that is expected
based on theory and observation of LSS, and its dependence on
the redshifts of the lens and source. 

The tensor components are random variables of zero mean, with variances
and covariances given in terms of the power spectrum of $\phi$.
For example, if $\tau_{1} \ge \tau_{2} \ge \tau_{3}$, then
following the method of Kaiser (1992) we find that
\begin{eqnarray}
\label{frms}
\langle {\rm F}_{ij}(\tau_{1},\tau_{2})\ {\rm F}_{kl}(\tau_{1},\tau_{3})
\rangle = 2 \pi^{2} {\rm Q}_{ijkl} \int_{0}^{\infty} k^{5} dk 
\int_{\tau_{2}}^{\tau_{1}} \frac{(\tau-\tau_{2})(\tau_{1}-\tau)}
{\tau_{1}-\tau_{2}} \frac{(\tau-\tau_{3})(\tau_{1}-\tau)}
{\tau_{1}-\tau_{3}} 
 P_{\phi}(k,\tau) d\tau\ ,
\end{eqnarray}
\[ \mbox{where Q}_{ijkl}= \left\{
\begin{array}{ll}
3 & \mbox{if $ijkl$ are all equal.} \\
1 & \mbox{if of $ijkl$ two $=x$ and two $=y$.} \\ 
0 & \mbox{otherwise.} 
\end{array} \right. \]
This assumes that the dominant contribution comes from modes with wavelengths 
that are much smaller than the distance $\tau_{1}-\tau_{2}$. This is 
satisfied for standard forms of the power spectrum and relevant distances.

\begin{figure}[t]
\vspace*{8.7 cm}
\caption{The top plot shows the r.m.s.\ value of $\frac{1}{2}$ Tr 
F$_{\eff}$, as a function of $z_{S}$, with $z_{L}$ set so that
$\rL=\frac{1}{2}\rS$.  The bottom right plot shows the same quantity,
but as a function of $z_{L}$, for fixed $z_{S}=3$.
The bottom left plot shows the r.m.s.\ value of 
$\frac{1}{2}$ Tr F$_{OL}$ as a function of $z_{L}$.
All curves use the non-linear power spectrum,
with $\Omega_{m}=1$, h$=0.25$, and $\sigma_{8}=0.8$.}
   \includegraphics{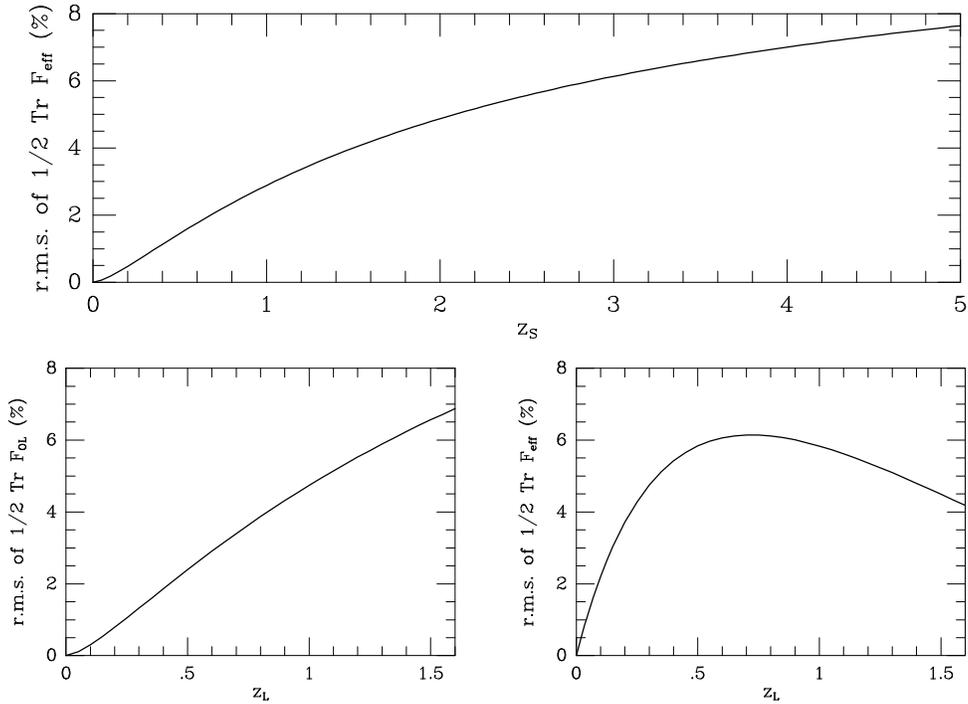}
\label{fig1}
\end{figure}

We follow the approach of Seljak (1995) 
for calculating r.m.s.\ shear. For the linear power spectrum we take a 
scale-invariant spectrum with a CDM type transfer function 
(\cite{BBKS}), which is normalized by $\sigma_{8}$, the mass
fluctuation averaged over a sphere of radius $8h^{-1}$ Mpc, and
whose peak is determined by $\Omega_{m0} h$. Galaxy and cluster
surveys are consistent with $\sigma_{8} \approx 0.8$ and 
$\Omega_{m0} h \approx 0.25$ (e.g.\ \cite{peacock}).
We then find the non-linear power spectrum using
the mapping proposed by Hamilton et al.\ (1991) and
extended by Peacock \& Dodds (1994), in the improved
form of Jain et al.\ (1995), which they show agrees 
with N-body simulations at the relevant scales, for an $\Omega_{m}=1$
universe. We find that the dominant contribution in equation (\ref{frms}) 
comes from wavenumbers $k \approx 3$ h Mpc$^{-1}$, with a
broad range of two decades on each side contributing significantly.
We therefore require a power spectrum that is accurate well into
the non-linear regime. 

The shear due to F$_{OS}$ is defined as $\Gamma=\sqrt{(\Gamma_{1})^{2}
+(\Gamma_{2})^{2}}$, where $\Gamma_{1}=\frac{1}{2}({\rm F}_{OS}^{11}
-{\rm F}_{OS}^{22})$ and $\Gamma_{2}={\rm F}_{OS}^{12}$. 
Equation (\ref{frms}) shows that $\Gamma$ has the same r.m.s.\ 
value as $\frac{1}{2}$ Tr F$_{OS}$, which is the convergence
or surface mass density $\kappa$ due to F$_{OS}$. The same
is true for F$_{\eff}$. Because each of the tensors F$_{OL}$ and F$_{LS}$
is correlated (and so tends to be aligned) with F$_{OS}$, 
F$_{\eff}$ tends to have smaller components than F$_{OS}$. 
For a given $z_{S}$, the r.m.s.\
shear of F$_{\eff}$ is maximized at $\approx 60 \%$ of that of F$_{OS}$,
approximately at $z_{L}$ for which $\rL=\frac{1}{2}\rS$. In the top plot 
of figure \ref{fig1} we show 
the r.m.s.\ value of $\frac{1}{2}$ Tr F$_{\eff}$ as a function of $z_{S}$, 
at this maximizing $z_{L}$. In the bottom right plot we show the same 
quantity, but as a function of $z_{L}$, for a fixed source at
redshift $3$. This quantity can be estimated for other redshifts
through its scaling $\propto r_{L}r_{LS}/\sqrt{r_{S}}$, which is accurate
to better than $10 \%$. The bottom left plot shows the r.m.s.\ value of 
$\frac{1}{2}$ Tr F$_{OL}$ as a function of $z_{L}$. This
quantity scales approximately as $\propto r_{L}^{3/2}$, and also
equals $\frac{1}{2}\sigma_{H_0,2}$ as shown in \S 4.
All curves use the non-linear power spectrum, which gives r.m.s.\
shear larger than the linear spectrum by a factor of $\approx 2.5$.
Equation (\ref{frms}) gives a statistical tendency for 
perpendicular alignment between F$_{OL}$ and F$_{\eff}$,
which increases $\sigma_{H_0,1}$ relative to $\frac{1}{2}$ Tr F$_{\eff}$.
At $\zS=3$ and $\rL=\frac{1}{2}\rS$, $\sigma_{H_0,1}=11.7\%$, and  
it scales approximately as $\propto r_L \sqrt{r_S-r_L/3}$.
Since the effect of LSS accumulates over distance, we find that the
induced shears and time delay uncertainties are all
smaller at lower redshifts. For the 0957+561 redshifts ($z_L=.36,
z_S=1.41$), the r.m.s. $\frac{1}{2}$ Tr F$_{\eff}$ is $3.7\%$,
$\sigma_{H_0,1}=5.9\%$, and $\sigma_{H_0,2}=3.3\%$. In addition, note that 
the effects of LSS on lens reconstruction 
disappear as $z_{L} \rightarrow 0$, even if $z_{S}$ is large.

The r.m.s.\ shear increases with $\sigma_{8}$, in exact proportion for the
linear power spectrum, faster for the non-linear spectrum. The
r.m.s.\ shear also grows with $h$ (at fixed $\sigma_{8}$),
by $\approx 35 \%$ for $h=0.5$ compared to $h=0.25$. As an
additional example, tilted CDM (e.g.\ \cite{cen}) with $h=0.5,\ \sigma_8=
.6$, and power spectrum index $n=0.8$ lowers the shear by $\approx
30\%$ compared to Figure 2.
The r.m.s.\ shear can also be calculated for models with $\Omega_{m0} 
\neq 1$ with modified formulas (see Appendix A).

\section{Illustration of the effect of LSS}

Kovner (1987) analyzed in some generality the properties of
the lens mapping for an axi-symmetric lens perturbed by a
weak shear. We simply wish to illustrate the possible
observable effects of a shear of the magnitude that
we obtained in \S 5.
We choose a particular symmetric lens distribution, a
singular isothermal sphere, with deflection law $\aV(r_{L}
\thV)=\thV / \theta$.
We use equations (\ref{lens}) -- (\ref{eqmag}) to find
the caustics and critical curves of the lens system.
The critical curves are the points in the image
plane for which the magnification det$^{-1}(\partial\beta^{i}/\partial\theta_{j})$ 
is infinite, and the caustics are 
the corresponding points in the source plane. The
caustics also determine image multiplicities, in that
a source located outside all the caustics has a single
image, and each time a source moves inside a caustic
two images are added (except that for a singular surface density,
one image is captured in the core when multiple
images are produced). For a given source
position, we can thus find all image positions, 
magnifications, and also time delays with equation (\ref{tdelay1}).

The components F$_{OL}^{ij}$, etc.\ are random variables,
with covariances obtained as in \S 5 above. We choose $z_{L}=0.78$ 
and $z_{S}=3.0$, and take a particular example:
$${\rm F}_{OL}=\left( \begin{array}{rr} -3.87 & 0.50 \\
0.50 & -2.04 \end{array} \right) \%\ {\rm ,\ F}_{LS}=\left( \begin{array}
{rr} -0.70 & 3.68 \\ 3.68 & 2.20 \end{array} \right) \%\ {\rm ,\ F}_{\eff}=
\left( \begin{array}{rr} 6.65 & -6.56 \\
-6.56 & -0.31 \end{array} \right) \%\ . $$

\begin{figure}[t]
\vspace*{7.7 cm}
\caption{Caustics in the source plane (upper panels) and critical 
curves in the image plane (lower panels)
for a singular isothermal sphere, with no LSS, and with LSS.
For the latter case, the $\vec{Y}$ (distorted source) plane and $\vec{X}$
(lens) plane are also shown. Also plotted are two 
source positions (marked $+$ and $\times$) and the corresponding images 
for each. A dot shows the $\thV=\bV=0$ position.}
   \includegraphics{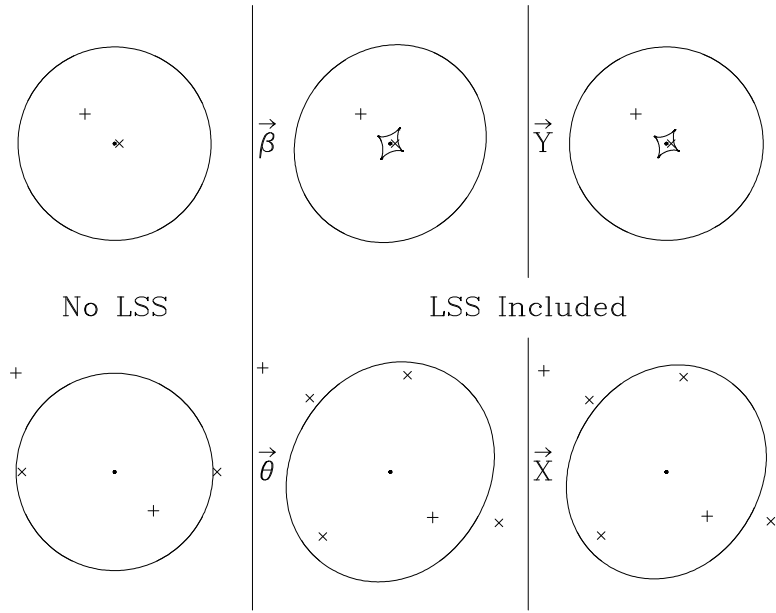}
\label{figlens}
\end{figure}

Figure \ref{figlens} shows the caustics in the source plane 
(upper panels) and critical curves in the image plane (lower panels), 
for the lens alone and for the lens plus LSS. For the latter
case, the $\vec{Y}$ (distorted source) plane and $\vec{X}$
(lens) plane are also shown. Also plotted are two source positions 
and the corresponding image positions. Table 1
lists the image positions, magnifications, and
relative time delays. LSS changes the image configurations significantly. 
It displaces images from the line to the lens, 
in the two-image configuration, and also produces four-image systems 
when $|\bV|$ is small. 

\begin{figure}[t]
\begin{center}
\begin{tabular}{||l|r|r|r|r||} \hline 
\multicolumn{5}{||c||}{Source at $\bV=(-0.30,0.30),\ 
\vec{Y}=(-0.31,0.30)$} \\ \hline  & Image Plane $(\thV)$ & 
Lens Plane $(\vec{X})$ & Magnification & Relative $\Delta$t \\ \hline
No LSS & $(-1.00,1.00)$ & --- & $3.36$ & --- \\ \cline{2-4}
 & $(0.39,-0.39)$ & --- & $-1.36$ & 0.84 \\ \hline
With LSS & $(-1.30,1.05)$ & $(-1.24,1.03)$ & $2.82$ & --- \\ \cline{2-4}
 & $(0.43,-0.46)$ & $(0.41,-0.45)$ & $-2.05$ & 0.98 \\ \hline \hline
\multicolumn{5}{||c||}{Source at $\bV=\vec{Y}=(0.05,0)$} \\ \hline
No LSS & $(1.04,0)$ & --- & $21.0$ & --- \\ \cline{2-4}
 & $(-0.94,0)$ & --- & $-19.0$ & 0.099 \\ \hline
With LSS & $(1.10,-0.52)$ & $(1.06,-0.50)$ & $6.18$ & --- \\ \cline{2-4}
 & $(-0.82,0.75)$ & $(-0.78,0.73)$ & $11.2$ & 0.098 \\ \cline{2-4}
 & $(0.18,0.98)$ & $(0.17,0.96)$ & $-9.22$ & 0.121 \\ \cline{2-4}
 & $(-0.68,-0.66)$ & $(-0.66,-0.65)$ & $-5.85$ & 0.165 \\ \hline
\end{tabular}
\end{center}

Table 1: Positions of the images shown in figure 3. 
Also listed are the absolute magnifications
(with a sign giving the image parity), and the time delay in units
of $r_{L}r_{S}/r_{LS}$ relative
to the earliest image to arrive at the observer.
\label{tablens}
\end{figure}

\section{Weak Lensing and Strong Lensing}

The approximation of equation (\ref{phi}) suffices for consideration of
strong lensing, where $| \thV |$ is very small ($\approx$ a few
arcseconds). In weak lensing, however, the shear is observed
at larger angles (arcminutes), and the variation of LSS
shear with angle is important. Moreover, as noted above,
there are potential convergence problems with our expansion
of $\phi$, even in the strong lensing case.
Our formalism allows us to make a more rigorous and powerful expansion,
and to study the transition from strong to weak lensing. Note that
for shape distortions we must use $\bV$ and $\thV$ rather than 
$\vec{Y}$ or $\vec{X}$, since we are interested in the observed compared 
to the intrinsic ellipticity of background galaxies.

We replace equation (\ref{phi}) with
\begin{equation}
\phi(r \hat{z} + \xP) = \sum_{n=0}^{\infty} \frac{1}{n!} 
(\xP \cdot \dP)^{n} \phi(r \hat{z})\ ,
\label{phiexp}
\end{equation}
where this expression must be consistent with equations
(\ref{eq1}) and (\ref{eq4}) for $\xP$. We now include derivatives to 
all orders in equation (\ref{phiexp}), but only keep terms linear
in $\phi$. This means that in the following expansions in
powers of $\thV$ and $\vec{\gamma}$, we are finding each
coefficient up to relative corrections of the same order as the
r.m.s.\ LSS shear.

Between the observer and the lens, equation (\ref{eq1}) requires
\begin{equation}
\label{eqrec}
\xP(\tau)=-(\tO-\tau)\nPO -2\int_{\tau}^{\tau_{0}}
(\tp-\tau)e^{\xP(\tau ') \cdot \dP}\dP\phi(\tp)d\tp\ ,
\end{equation}
where the exponential denotes the corresponding Taylor
series expansion (and $\phi$ on the RHS is again evaluated 
on the $z$-axis). We can find all terms linear in $\phi$ in the
solution by substituting for $\xP(\tp)$ in
the RHS the $\phi$-independent term $-(\tO-\tp)\nPO$.
We now find that 
\begin{eqnarray}
l_{\perp}^{i}(\tau)&=&-\theta^{i}+2\int_{\tau}^{\tau_{0}}
\left[ e^{(\tau_{0}-\tau ')\vec{\theta} \cdot \dP}-1 \right]
\partial^{i}\phi(\tp) d\tp
\ , \nonumber \\
d_{\perp}^{i}(\tau)&=&(\tO-\tau)\theta^{i}-
2\int_{\tau}^{\tau_{0}} (\tp-\tau)
\left[ e^{(\tau_{0}-\tau ')\vec{\theta} \cdot \dP}-1 \right]
\partial^{i}\phi(\tp) d\tp\ .
\label{weakeq1}
\end{eqnarray}
If we let $\pP=\lP(\tL)+\gV(\dPL)$ then between the
lens and the source we similarly find that
\begin{eqnarray}
l_{\perp}^{i}(\tau)&=&p_{\perp}^{i}+2\int_{\tau}^{\tau_{L}}
\left[ e^{(\dPL-(\tau_{L}-\tau ')\vec{p}_{\perp}) 
\cdot \dP}-1 \right]
 \partial^{i}\phi(\tp) d\tp
\ , \nonumber \\
d_{\perp}^{i}(\tau)&=&d_{\perp}^{i}(\tL)-(\tL-\tau)p_{\perp}^{i}-
2\int_{\tau}^{\tau_{L}} (\tp-\tau)
\left[ e^{(\dPL-(\tau_{L}-\tau ')\vec{p}_{\perp}) 
\cdot \dP}-1 \right]
\partial^{i}\phi(\tp) d\tp\ . 
\end{eqnarray}

The lens equation is
\begin{eqnarray}
\beta^{i} &=& \theta^{i}-\alpha^{i}- 
 \frac{2}{r_{S}} \int_{\tau_{L}}^{\tau_{0}} (\tau-\tau_{S})
\left[ e^{(\tau_{0}-\tau)\vec{\theta} \cdot \dP}-1 \right]
\partial^{i}\phi(\tau) d\tau \nonumber \\
&& - \frac{2}{r_{S}} \int_{\tau_{S}}^{\tau_{L}} (\tau-\tau_{S})
\left[ e^{((\tau_{0}-\tau)\vec{\theta} - (\tau_{L}-\tau)\vec{\gamma})
\cdot \dP}-1 \right] \partial^{i}\phi(\tau) d\tau\ ,
\label{lensbig}
\end{eqnarray}
with $\aV$ and $\vec{\gamma}$ evaluated at $\dPL$ calculated
from equation (\ref{weakeq1}).
The magnification matrix is 
\begin{equation}
\frac{\partial\beta^{i}}{\partial\theta_{j}}=\delta^{ij}
+{\rm A}^{ij}-\Psi^{ij}-{\rm B}^{ij}\ ,
\label{magbig}
\end{equation}
where 
\begin{eqnarray}
{\rm A}_{ij}&=&-\frac{2}{r_{S}} \int_{\tau_{L}}^{\tau_{0}} 
e^{(\tau_{0}-\tau)\vec{\theta} \cdot \dP} (\tau-\tau_{S})
(\tO-\tau) \partial_{i}\partial_{j}\phi(\tau) d\tau \nonumber \\
&& - \frac{2}{r_{S}} \int_{\tau_{S}}^{\tau_{L}} 
e^{((\tau_{0}-\tau)\vec{\theta} - (\tau_{L}-\tau)\vec{\gamma})
\cdot \dP} (\tau-\tau_{S}) (\tO-\tau)
\partial_{i}\partial_{j}\phi(\tau) d\tau\ , \nonumber \\
{\rm B}_{ij}&=&-\frac{2}{r_{L}} \Psi^{k}_{i} \int_{\tau_{L}}^{\tau_{0}} 
e^{(\tau_{0}-\tau)\vec{\theta} \cdot \dP}
(\tau-\tau_{L}) (\tO-\tau)
\partial_{k}\partial_{j}\phi(\tau) d\tau \nonumber \\
&& - \frac{2}{r_{LS}} \Psi^{k}_{j}\int_{\tau_{S}}^{\tau_{L}} 
e^{((\tau_{0}-\tau)\vec{\theta} - (\tau_{L}-\tau)\vec{\gamma})
\cdot \dP} (\tau-\tau_{S}) (\tL-\tau)
\partial_{k}\partial_{i}\phi(\tau) d\tau\ . \nonumber
\end{eqnarray}

If we expand the exponentials to first order in equation (\ref{lensbig})
and zeroth order in equation (\ref{magbig}), we recover the 
results of \S 3. If lensing is not strong,
B$_{ij}$ is small, and in the external shear A$_{ij}$
we can set $\vec{\gamma}=(\thV-\bV)r_{S}/r_{LS}$. In
the limit of weak lensing, we can set $\vec{\gamma}=0$ to get 
\begin{equation}
{\rm A}_{ij}=-\frac{2}{r_{S}} \int_{\tau_{S}}^{\tau_{0}} 
e^{(\tau_{0}-\tau)\vec{\theta} \cdot \dP} (\tau-\tau_{S})
(\tO-\tau) \partial_{i}\partial_{j}\phi(\tau) d\tau\ .
\label{appB}
\end{equation}
This expression can be used to calculate two point correlation functions
of ellipticity. E.g., we can write down $\left <{\rm Tr A}(\theta=0){\rm Tr A}
(\theta) \right >$ and evaluate this expectation value in Fourier space. The 
exponential of $ir\thV\cdot\kV$ (in Fourier space) oscillates rapidly at high $k$, 
which cuts off small wavelengths and prevents any divergence. The result, which is 
derived fully in Appendix B, agrees with previous 
analyses of weak lensing in the absence of a primary lens (e.g.\ \cite{blandford};
\cite{miralda}; \cite{kaiser}). These analyses have found that
the relative change in the angular correlation of ellipticity is
smaller than $10\%$ (in an r.m.s.\ sense) for $\theta$ less than about an
arcminute. For the non-linear spectrum, we find this to be true below
about $10\arcsec$ (see also \cite{seljak95}), thus justifying our keeping 
only linear terms in $\thV$ for strong lensing. Our result (\ref{magbig})
is more general than weak lensing, as it includes a primary lens ($\Psi^{ij}$)
and cross-terms ($B^{ij}$).

We can also get quadratic and higher-order terms in the gradients
of $\phi$ by iterating this procedure. Given a
solution $\xP^{(j)}$ we substitute it in the RHS of equation 
(\ref{eqrec}) and find the next order solution $\xP^{(j+1)}$.
The exponential of $i\vec{k}\cdot\xP^{(j)}$ ensures that
small wavelength modes are cut off in the calculation
of $\xP^{(j+1)}$. This corresponds to the physical intuition that on average 
$\phi(r \hat{z} + \xP^{(j)})-\phi(r \hat{z})$ is determined by power on scales 
of order $|\xP^{(j)}|$. If we calculate the r.m.s.\ shear at a point 
(i.e.\ for $\theta=\gamma=0$) corresponding to $\xP^{(j+1)}$, for a given 
$\xP^{(j)}(\tau)$, the 
answer is the same as for $\xP^{(j)}(\tau)=0$ as long as the angular deflection
is small and LSS power on the scale of $r_{S}$ is negligible. There is, of course a 
statistical correlation between
$\xP^{(j)}(\tau)$ and $\phi(\tau)$, but it is typically weak since 
$\xP^{(j)}(\tau)$ is determined by the accumulated deflection from
$\tau$ to $\tO$, a distance many times larger than the coherence length
of LSS. The first correction to $F_{OS}$ is a relative correction of order 
$1\%$, if $\phi$ is Gaussian, and the corrections to $F_{OS}$ are expected to be 
small also for a non-Gaussian $\phi$ produced through hierarchical clustering.

\section{Conclusions}

We have shown that LSS can have significant effects on strong 
gravitational lensing. This suggests that lens reconstruction
done without including LSS might reach incorrect conclusions
about the distribution of matter in the lensing galaxy or
cluster. It also raises the possibility of
constraining the amplitude of the power spectrum directly, 
if lensing observations can be used to detect the effect of LSS. 

The effect of LSS is simply described by two symmetric
tensors. Including only the effect of F$_{\eff}$, we find 
that the observed power spectrum of LSS requires
the presence of an external shear of order $6\%$ if 
$z_{S}=3$. This can significantly affect the cross 
sections for image multiplicities in lens systems. In particular,
it can produce more images than would be created in the
absence of LSS. This implies that in addition
to the usual magnification bias, which increases the
observed number of quads relative to doubles, there is a bias in 
quads toward lines of sight with relatively large effective shear
from LSS.

The second effect, given by F$_{OL}$, produces a magnification and
shear between the observer and the lens. This term enters the lens
equation differently from the effective shear and should be
included in lens modelling. It also distorts the lens plane, which 
contributes an ellipticity to the observed lens galaxy, and converts 
the angular diameter ``distance'' into a tensor, though the comoving 
distance is still simply defined in terms of the observed redshift. 
Even if lens reconstruction can model the lens potential exactly,
we find that LSS creates an absolute uncertainty ($\approx 5-10 \%$ 
at $1\sigma$) in deductions of the Hubble constant from time delays.
Among lens systems, the uncertainty is smaller for those with lower lens 
and source redshifts.

Models of quadruple lenses typically
find a shear of order $10\%$ in addition to an
axi-symmetric mass distribution. If
this is due to the ellipticity of the lens galaxy,
it may imply a larger ellipticity than that observed
in the galaxy light distribution, as confirmed in a number
of cases by recent observations. 
Since quads tend to be produced more easily when the shear 
due to the galaxy and the effective shear due to LSS are aligned, it 
is important to compare the magnitudes of the observed
and modeled shears for consistency, and not only their directions.
If the shear is due instead to other galaxies or clusters near the line of 
sight to the source, these additional lenses may not be found
where expected. Only high-resolution observations and
careful modelling of particular lens systems will 
determine if the shear may in part be due to LSS. When the parameters
of many lenses are confidently known, it may become possible to
study the redshift dependence of the shear. E.g., LSS does not
affect lens reconstruction if the lens is at very low redshift.
The original Einstein Cross 2237+0305 has $z_{L}=0.04,\ z_{S}=1.7$, and
the lens light distribution seems to be consistent with the lensing mass
(Rix et al.\ 1992). Other methods must be used to investigate independently 
whether the mass in galaxies is more flattened than the light 
distribution or not. E.g., an affirmative answer is suggested by an
optical plus X-ray study of the elliptical galaxy NGC 720 (\cite{buote}).

Constraining the effects of LSS on strong lensing should
complement observations of weak lensing due to LSS.
For measurements of weak lensing the sources are background
galaxies, and the interpretation is complicated by the
unknown source redshift distribution, while for some
strong lenses the redshifts of the lens and source
are known. If the characteristic source redshift for weak lensing is 
$\approx 0.7 - 1$ then the shear due to LSS is significantly
smaller than for strong lensing (e.g.\ figure \ref{fig1}). 
In addition, since measurements of weak lensing with high signal 
to noise require relativly large angular fields, the r.m.s.\ 
shear is further reduced. On the other hand, weak
lensing due to LSS can in principle be distinguished from 
other effects by averaging over a wide field, an option
not available in strong lensing. Demanding consistency
between determinations of the effects in these two
regimes should allow us to learn more about the
distribution of matter in the universe.

\acknowledgements
I thank Ed Bertschinger for suggesting this problem and for helpful
advice, Uro\v{s} Seljak for valuable discussions and for
his computer program to calculate r.m.s.\ shear, 
Paul Schechter for helpful discussions and comments, and the referee Josh 
Frieman for valuable comments. This work was supported by NASA grant NAG5-2816.
\appendix
\section{Appendix A}

To calculate LSS shear in a curved background requires
slight modifications of our formulas (e.g.\ \cite{miralda}; 
\cite{seljak95}).
In a general Robertson-Walker model, the line element is 
\begin{equation}
ds^2=a^2(\tau)\biggl[-(1+2\phi)d\tau^2+(1-2\phi)
[d\chi^2+\sin_K^2\chi(d\theta^2+\sin^2 \theta d\phi^2)]
\biggl]\ ,
\end{equation}
in terms of spherical comoving coordinates, where we
are now using the variable $\chi=\tau_{0}-\tau$. We have
defined
\begin{eqnarray}
\sin_K\chi \equiv
\left\{ \begin{array}{ll} K^{-1/2}\sin K^{1/2}\chi & 
\mbox{if } K>0.\\
\chi & \mbox{if } K=0.\\
(-K)^{-1/2}\sinh (-K)^{1/2}\chi & \mbox{if } K<0.\\
\end{array}
\right.
\end{eqnarray}
The curvature is $K=(\Omega_0-1)H_0^2$. The relation between
redshift and $\tau$ is given by 
the Friedmann equation. 

In a curved geometry, a deflection by angle $\delta \thV$
at $\chi '$ leads to a perpendicular displacement at $\chi$ of
$\delta \xP=\delta \thV \sin_K(\chi-\chi ')$.
In our approximation of \S 3, these deflections simply
add linearly. Thus, our expressions for $\xP$ or $\ddP$ remain
valid if we replace any expression of the form $\tau_{1}-
\tau_{2}$ with $\sin_K(\tau_{1}-\tau_{2})$, so e.g.\ $r_{LS}=
\sin_K(\tau_{L}-\tau_{S})$.
Thus the lens equation (\ref{lens}), the magnification matrix
(\ref{eqmag}), and (again by Fermat's principle) the time 
delay (\ref{tdelay1}) all have the same form except that now
\begin{equation}
{\rm F}_{ij}(\tau_{1},\tau_{2}) = -\frac{2}{\sin_K(\tau_{1}-
\tau_{2})}\int
_{\tau_{2}}^{\tau_{1}}\sin_K(\tau-\tau_{2}) 
\sin_K(\tau_{1}-\tau)\partial_{i}\partial_{j}\phi(\tau)d\tau\ .
\end{equation}

\section{Appendix B}

In the limit of weak lensing with no primary lens, our formalism
reproduces previously derived results. From equation (\ref{appB}), we
find $$\left <{\rm Tr A}(\theta=0){\rm Tr A}(\theta)\right > 
=\frac{4}{r_S^2}\int_0^{r_S}dr_1 \int_0^{r_S}dr_2\ e^{r_1 \vec{\theta} 
\cdot \dP} r_1 (r_S - r_1) r_2 (r_S - r_2) \left < \dP^2
\phi(\tau_1) \dP^2 \phi(\tau_2) \right >\ ,$$
where we have used $r_1=\tau_0 - \tau_1$, etc. We convert to Fourier
space, and use spherical coordinates $\{k,\ \theta_k,\ \phi_k\}$.
We use the approximation that only $k$ values for which $kr_{S} \gg 1$
(i.e., wavelengths much smaller than the source distance)
make an important contribution. This implies that $\int_0^{r_S} dr_1
\int_0^{r_S} dr_2 \approx \int_0^{r_S} dr_1 \int_{-r_S}^{r_S} du$, with
$u=r_2 - r_1$, and also that we can set $r_2=r_1$ in the distance terms in the 
integrand. Letting $\omega=k u$ and denoting $r_1$ now by $r$, 
in Fourier space our expression 
becomes $$4 \int_0^{r_S} dr \int_{-k r_S}^{k r_S} d\omega \int d^3 k\ e^{i k r
\theta \sin\theta_k \cos\phi_k} e^{i \omega \cos\theta_k} \sin^4 \theta_k\ 
r^2 \left(1-\frac{r}{r_S}\right)^2 k^3 P_{\phi}(k,\tau=\tau_0-r)\ ,$$
where in the $\vec{k}$ integration we chose the $x$-axis in the direction
of $\thV$. Under the approximation of $k r_S \gg 1$, $$\int_{-k r_S}^{k r_S}
d \omega e^{i \omega \cos\theta_k}=2\pi \delta(\cos\theta_k)\ .$$
Our expression thus equals $$8 \pi \int_0^{r_S} dr\ r^2 \left(1-\frac{r}{r_S}
\right)^2 \int_0^{\infty} k^5 dk P_{\phi}(k,\tau=\tau_0-r) \int_0^{2 \pi}
d\phi_k\ e^{i k r \theta \cos\phi_k}\ ,$$ or, finally, 
\begin{equation}
\left <{\rm Tr A}(\theta=0){\rm Tr A}(\theta)\right > = 16 \pi^2 \int_0^{r_S} 
dr\ r^2 \left(1-\frac{r}{r_S} \right)^2 \int_0^{\infty} k^5 dk P_{\phi}(k,
\tau=\tau_0-r) J_0(k r \theta)\ .
\end{equation}
This correlation function of Tr A equals that of twice the shear of A, which has 
also been derived previously through other methods 
(e.g.\ \cite{blandford}; \cite{miralda}; \cite{kaiser}).

\end{document}